\documentclass[aps,preprint,amsmath,amsfonts,amssymb,showpacs,floatfix]{revtex4}

\usepackage{graphicx}

\begin{document}

\title{Crumpling of Curved Sheets: Generalizing F\"{o}ppl-von 
K\'arm\'an}

\author{J. Leo van~Hemmen$^1$ and Mark A. Peterson$^2$}

\affiliation{$^1$Physik Department, Technische Universit\"at
M\"unchen , 85747 Garching bei M\"unchen, Germany.}
\affiliation{$^2$Physics Department, Mount Holyoke College, South
Hadley, MA 01075-6420}


\begin{abstract}
We generalize the F\"{o}ppl-von K\'arm\'an equations to an initially 
precurved sheet and present the underlying derivation.  A 
geometrically computed moment of strain replaces the notion of bending 
moment and results in a geometric formulation of the theory of shells.  
As the curvature approaches zero, i.e., the sheet becomes flat, the 
new equations reduce to the classic F\"{o}ppl-von K\'arm\'an ones.  
The present theory solves the long-standing problem of formulating 
these equations for an \emph{a priori} curved shell and applies, for 
instance, both to shell theory and to strongly curved biomembranes of 
cells as closed surfaces, exhibiting crumpling as the membrane 
thickness goes to zero.
\end{abstract}

\maketitle

\begin{figure}[b]
\includegraphics[width=0.89\columnwidth]{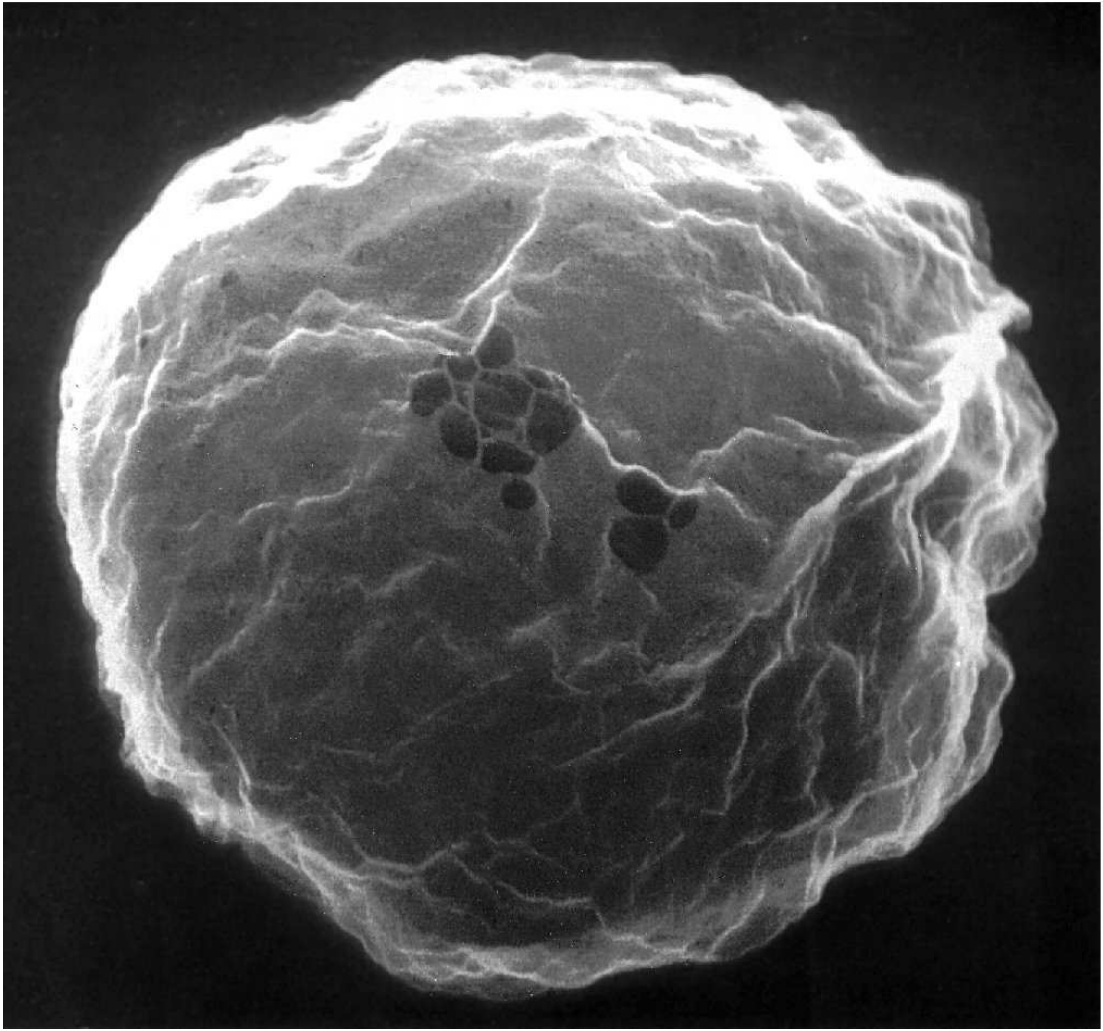}
\caption{Scanning electron micrograph \cite{steck} of a `ghost', a red
blood cell (erythrocyte) voided of its cytoplasm through a small hole
created by osmotic rupture of the cell (perhaps the dark area in the
membrane).  It consists of a closed cell membrane (plus reticulum) of
thickness $h \approx 10 \,$nm; the cell's diameter is about 10~$\mu$m.
The ridges give the membrane a crumpled structure.  As was suggested
by Lobkovsky \cite{lobkovsky}, one can understand this ``chaotic''
structure through a system of two coupled nonlinear partial
differential equations due to F\"{o}ppl \cite{foeppl} and von
K\'arm\'an \cite{vonK} (FvK).  One of the two differential operators
of the form $\Delta^{2}$ has a small prefactor scaling as $h^2$ where
$h$ is the membrane thickness; cf.  (\ref{CellEqNormal2}) and
(\ref{FvK2}) below.  In the limit $h \to 0$ it cannot be neglected but
gives rise to crumpling -- analogous to $R_{e}^{-1} \Delta$ in
hydrodynamics where it generates turbulence as the Reynolds number
$R_{e} \to \infty$.  The FvK equations have been derived as a local
deviation from the flat state.  The cell membrane, however, was never
flat and clearly shows that in biological physics this assumption
often has no justification.  Picture courtesy of T.L. Steck.  The
original has been contrast enhanced.}
\label{crumpRBC}
\end{figure}

Crumpling of thin membranes or sheets and the existence of auxetic
(negative Poisson ratio) materials have challenged physical
imagination, and explanation, for quite a while
\cite{np87,sn88,lglmw95,lobkovsky,lw97,bap97,kramer,
ccmm99,audoly99,bpcba00,mwwn02,ddwvk02,bowick,lidmar}.
Figure~\ref{crumpRBC} shows a striking example stemming from the
biophysics of membranes \cite{steck,es94,npw04}.  At present it
appears that all these phenomena occurring in thin sheets are
realizations of the F\"{o}ppl-von K\'arm\'an (FvK) equations
\cite{foeppl,vonK,ll59,ehm64}, an important case of the equilibrium
equations of elasticity theory, incorporating nonlinearity, and aiming 
at describing strongly curved sheets of finite thickness $h>0$.  The 
FvK equations only apply, however, to a thin sheet that is planar to 
start with and is then bent strongly.  Their derivation could not 
handle a sheet being \emph{pre}curved, although this often occurs in 
practice -- as in Fig.~1, for instance.  The latter problem has been 
tantalizing the physics of membranes ever since F\"{o}ppl
\cite{foeppl} and von K\'arm\'an \cite{vonK} proposed their equations
a century ago.

The FvK derivation includes the assumption that the sheet in question
is thin, but it allows for the possibility that deviations from the
initial shape may be large: it goes to quadratic order in the normal
deformation parameter.  Because the sheet is thin, by assumption, it
is still consistent to assume a linear stress-strain relationship
within the sheet.  Several studies have shown \cite{ehm64} that this
description makes physical sense for large deformation, up to and
including sharp creasing -- a regime well beyond what F\"{o}ppl and
von K\'arm\'an originally had in mind.

There is yet another assumption in FvK, which greatly limits its
applicability to physical systems of interest, namely, that the
initial shape of the sheet to be strongly bent or even crumpled is
planar.  How would deformation of an initially \emph{curved},
anisotropic surface interact with the pre-existing anisotropy?  This
and other fundamental questions of strong bending and crumpling of
surfaces cannot even be asked in the context of the original FvK
theory.  We present here the generalization necessary to handle such
questions.

If one looks to shell theory, a notoriously difficult part of
elasticity theory that deals with thin elastic surfaces
\cite{fluegge}, one finds that existing formulations are inadequate
for the purpose of handling \emph{a priori} curved sheets.  Shell
theory was developed in the context of mechanical engineering with
practical applications in mind.  The vocabulary of shell theory is
entirely physical.  One would not know from reading shell theory that
there is already a highly developed mathematical language for dealing
with surfaces and their deformations, namely differential geometry.
In this paper we show how to use methods of differential geometry to
formulate, and physically interpret, the equations of crumpling for a
general sheet.  It will be evident that this generality is necessary
in describing deformations in which the change in curvature is
comparable to pre-existing curvature.  By way of motivation we call
attention to a typical, biological, example in Fig.~\ref{crumpRBC}, an
initially curved surface that has subsequently been wrinkled.

We consider the equilibrium state of an elastic body occupying a
region $D \subset \Bbb{R}^{3}$ and subject to a conservative force
density $f$ derivable from a potential $U$.  The body is described by
an elastic energy density $E_{\rm el}$ quadratic in strain $u_{ij}$.
Thus its equilibrium is characterized by
\begin{equation}
    0=\delta \int_D {\rm d}V \,(E_{\rm el}+U)
    =\int_D {\rm d}V \,(\sigma^{ij}\delta u_{ij}-f^i\delta  u_i)
    \label{equilib}
\end{equation}
where the stress $\sigma^{ij}$ is symmetric and linear in strain, and 
$\delta u_i$ is an arbitrary small deformation, with $\delta u_{ij}$ a 
function of $\delta u_i$.  Now we ask what this means in case the body 
is a thin sheet, meaning that the region $D$ is a smooth surface 
$\mathcal{M}$ thickened along the normal direction by a small amount 
$h$, with $\mathcal{M}$ as the midsection.  (Generalizations to such 
cases as $h$ variable rather than constant, elastic constants 
dependent on position, etc., are straightforward, and will not be 
emphasized in what follows.) Let the Euclidean metric of three 
dimensional space be $g_{ij}$, and let the first, second, and third 
fundamental forms of $\mathcal{M}$ be $g_{\alpha\beta}$, 
$h_{\alpha\beta}$, $k_{\alpha\beta}$, where Greek indices take values 
$1,2$.  In particular the first fundamental form $g_{\alpha\beta}$ is 
just $g_{ij}$ restricted to $\mathcal{M}$.  We assume $\mathcal{M}$ is 
oriented by choice of a unit normal {\bf n}.

As is well known \cite{hilbert}, there is an orthogonal coordinate
system $(q^1,q^2)$ on $\mathcal{M}$ such that the level lines of the
coordinates are tangent to the principal curvature directions at every
point (umbilic points may be coordinate singularities).  Furthermore,
in a thickened three-dimensional neighborhood of $\mathcal{M}$ this
coordinate system may be extended to an orthogonal system
$(q^1,q^2,z)$, where the third coordinate $z$ is along the normal to
$\mathcal{M}$ through $(q^1,q^2)$, positive in the direction of {\bf
n}.  Methods for computing geometrical objects within this framework
have been given in \cite{peterson}.  In particular, in these
coordinates the metric $g_{ij}$ in $D_\mathcal{M}$ takes the form
\begin{equation}
   g_{ij}={\rm diag} \, (g_{11}(1-\kappa_1z)^2, g_{22}
   (1-\kappa_2z)^2,1) \ .
\end{equation}
Here ${\rm diag}(g_{11},g_{22})$ is the first fundamental form of
$\mathcal{M}$ and $(\kappa_1,\kappa_2)$ are the principal curvatures
of $\mathcal{M}$, functions of $(q^1,q^2)$ but not $z$.  The minus
signs define the sign conventions relating $(\kappa_1,\kappa_2)$ and
{\bf n}.  For example, if {\bf n} is the inner normal on the sphere,
then the principal curvatures are positive.  There is still the
freedom to choose $(q^1,q^2)$ such that at a given point $P\in
\mathcal{M}$ the first fundamental form obeys $g_{11,1}=g_{22,2}=0$
(at $P$).  Throughout what follows $_{,\alpha}$ denotes a partial
derivative $\partial/\partial x_{\alpha}$ with respect to the
coordinate $x_{\alpha}$ following the comma.  The only non-vanishing
Christoffel symbols at $P$ are now
\begin{eqnarray}
\Gamma^1_{12}&=&\Gamma^1_{21}=-\frac{1}{2}g^{11}g_{11,2}\\
\Gamma^1_{22}&=&\frac{1}{2}g^{11}g_{22,1}
\end{eqnarray}
and the corresponding ones with $1\leftrightarrow 2$.

A central ingredient in this approach is the observation that the two
dimensional strain tensor on $\mathcal{M}$ is related to the metric
tensor by
\begin{equation}
\delta u^\mathcal{M}_{\alpha\beta}=\frac{1}{2}\delta g_{\alpha\beta}
\end{equation}
where $\delta g_{\alpha\beta}$ is the change in the metric tensor
under a deformation $V=(\delta u^1,\delta u^2,\psi)$, a vector field.
Here the notation for the components of $V$ has been chosen to suggest
that tangential displacements $\delta u^\alpha$ are small, but that
the normal displacement $\psi$ is not necessarily small, reflecting
the anisotropy of e.g.\ a biomembrane.  The change in the metric
$g_{\alpha\beta}$ under $V$, in turn, can be computed as a Lie-Taylor
series, using the Lie derivative $\pounds_V$ \cite {schutz, frankel}
\begin{equation}
    \!\!       \delta g_{\alpha\beta}=(\pounds_V
   g)(\partial_\alpha,\partial_\beta) +\frac{1}{2}(\pounds_V(\pounds_V
   g))(\partial_\alpha,\partial_\beta) +\ldots
\end{equation}
Here $\pounds_V g$ is the Lie derivative of the three-dimensional
metric, with the result restricted to $\mathcal{M}$ by evaluating
on tangent vectors $\partial_\alpha$ to $\mathcal{M}$.  The result
depends only on the restriction of $V$ to $\mathcal{M}$, and if
$V$ is given only on $\mathcal{M}$, $V$ may be extended in any
smooth way for the purpose of the computation, for example the
components of $V$ could be independent of $z$.  We therefore
interpret $\delta u^\mathcal{M}_{\alpha\beta}$ as the strain in
$D_\mathcal{M}$, averaged across the thickness $h$.  The result
is, keeping terms linear in the deformation $\delta u^\alpha$, and
going to quadratic terms in $\psi$,
\begin{eqnarray}
   \delta u^\mathcal{M}_{\alpha\beta} &=& (\nabla_\alpha 
    \delta u_\beta + \nabla_\beta \delta u_\alpha +
   \nabla_\alpha\psi\nabla_\beta\psi+ \psi^2k_{\alpha\beta})/2
   \nonumber \\
   &&  -\psi h_{\alpha\beta}
  \label{uM}
\end{eqnarray}
where $\nabla_\alpha$ is the covariant derivative on $\mathcal{M}$.
(First derivatives of the function $\psi$ are, of course, just
ordinary partial derivatives.) We illustrate the method by computing
one component of the first Lie derivative.  The vector field $V$ is
the first order differential operator
\begin{equation}
V=\delta u^1\partial_1+\delta u^2\partial_2+\psi\partial_z \ .
\end{equation}
Then we have, evaluating at $P$ for simplicity,
and noting that $z=0$ there,
\begin{eqnarray*}
(\pounds_Vg)(\partial_1,\partial_1)&=&Vg(\partial_1,\partial_1)+
2g([V,\partial_1],\partial_1)\\
&=&\delta u^2g_{11,2}-2\psi \kappa_1g_{11}+2g_{11}\delta u^1_{,1}\\
&=&2(\delta u_{1,1}+\delta u_2\Gamma^2_{11})-2\psi g_{11} \kappa_1
\end{eqnarray*}
and similarly for other components.  Since the result is a tensor and 
coincides with $\nabla_\alpha \delta u_\beta+\nabla_\beta\delta 
u_\alpha-2\psi h_{\alpha\beta}$ in this coordinate system, it must be 
this tensor.  The second derivative computation is similar.

Since the three-dimensional body $D_\mathcal{M}$ is thin, we assume
the strain is at most linear in $z$.  But when $D_\mathcal{M}$ is
bent, it is relatively compressed on one side and extended on the
other, so it is clear that there \emph{is} a physically important
strain that is linear in $z$.  From the geometrical meaning of the
second fundamental form $h_{\alpha\beta}$ as the rate of rotation of
the unit normal {\bf n}, it is clear that this contribution is
$-z\delta h_{\alpha\beta}$, where $\delta h_{\alpha\beta}$ is the
change in the second fundamental form of $\mathcal{M}$ under the
deformation.

Unlike the first fundamental form, the second fundamental form is not
the restriction of a tensor field in three dimensions to
$\mathcal{M}$, and so its change under deformation must be computed in
a different way.  For example, the second fundamental form of the
varied surface $\mathcal{M}'$ is, up to a factor $-2$, the Lie
derivative of $g$ with respect to the unit normal on $\mathcal{M}'$,
according to (\ref{uM}), and there are also more classical ways to
compute it \cite{vhl}.  The result is
\begin{equation}
\label{uB}
   \delta u^B_{\alpha\beta}=-z\delta h_{\alpha\beta}=
   -z(\nabla_\alpha\nabla_\beta \psi - \psi k_{\alpha\beta}) \ .
\end{equation}
Here we have only gone to linear order in $\psi$, as the bending term
is usually small in any case, and the second order terms in $\psi$ are
complicated.  Again we illustrate the method.  Let the equation of the
varied surface $\mathcal{M}'$ be $z=\psi$, the result of deforming
$\mathcal{M}$ by the vector field $U=\psi\partial_z$.  Tangent vectors
to $\mathcal{M}'$ are
\begin{equation}
    X_\mu=\partial_\mu+[\partial_\mu,U]=\partial_\mu+\psi_{,\mu}
    \partial_z \ .
\end{equation}
The unit normal $\mathbf{n}'$ on $\mathcal{M}'$ is orthogonal to these
and normalized, hence to first order in $\psi$
\begin{equation}
\label{nprime}
   \mathbf{n}' = -g^{11} \psi_{,1} \partial_1-g^{22} \psi_{,2}
   \partial_2 +\partial_z \ .
\end{equation}
Then computing one component, at a point over $P$ for simplicity,
we have
\begin{eqnarray*}
   (\pounds_{\mathbf{n}'}g)(X_1,X_1) &=& \mathbf{n}' g(X_1,X_1)+
   2g([X_1,\mathbf{n}'],X_1) \nonumber\\
   &=&-\psi_{,2} \, g^{22}g_{11,2}
   +2g_{11}(1-z\kappa_1)(-\kappa_1) \nonumber\\
   & & -2\psi_{,11}~ \ .
\end{eqnarray*}
We evaluate at $z=\psi$, finding
\[
(\pounds_{\mathbf{n}'}g)(X_1,X_1)=-2g_{11}\kappa_1+2\psi g_{11}\kappa_1^2
-2(\psi_{,11}+\psi_{,2}\Gamma^2_{11})
\]
and similarly for other components, giving the result above. Going
to higher order in $\psi$ would only require solving to higher
order in (\ref{nprime}) as was done, for example, in
\cite{peterson}.

As stress is linear in strain, it too will be linear in $z$:
\begin{equation}
   \label{sigma}
   \sigma^{\alpha\beta}=\sigma_\mathcal{M}^{\alpha\beta}
   +z\sigma_B^{\alpha\beta} \ .
\end{equation}
Putting $\delta u_{\alpha\beta}=\delta u^\mathcal{M}_{\alpha\beta}
+\delta u^B_{\alpha\beta}$ from (\ref{uM}), (\ref{uB}), and
$\sigma^{\alpha\beta}$ from (\ref{sigma}) into (\ref{equilib}),
and integrating $z$ from $-h/2$ to $h/2$ we obtain
\[
   0=\int_\mathcal{M} {\rm d}A \sqrt{\sf{g}} \,[h
   \sigma_\mathcal{M}^{\alpha\beta}
   \delta u^\mathcal{M}_{\alpha\beta}
   +\frac{h^3}{12}\sigma_B^{\alpha\beta}\delta u^B_{\alpha\beta} -h
   (f^\alpha \delta u_\alpha +f^z \psi)]
\]
where ${\sf g} := \det (g_{\alpha\beta})$.
Finally, integrating by parts and ignoring boundary terms (if
$\mathcal{M}$ is a closed surface, for example), and recognizing that
the variations are arbitrary, we obtain the equations of equilibrium
\begin{eqnarray}
\label{CellEqTangential}
   0&=&\nabla_\beta\sigma_\mathcal{M}^{\alpha\beta}+f^\alpha \ ,\\
   0&=&\frac{h^3}{12}(\nabla_\alpha\nabla_\beta
   \sigma_B^{\alpha\beta}-k_{\alpha\beta}
   \sigma_B^{\alpha\beta})  \nonumber \\
   &+&h\sigma_\mathcal{M}^{\alpha\beta}(h_{\alpha\beta}
   +\nabla_\alpha\nabla_\beta\psi- k_{\alpha\beta}\psi)+h \left( f_z 
   -\nabla_\alpha\psi f^\alpha \right) \ ,   
   \label{CellEqNormal2}
\end{eqnarray}
a nonlinear system of equations for deformation $(u_\alpha,\psi)$ in
response to the force field $f$.

It is interesting to notice that one effect of keeping
\emph{second}-order terms in $\psi$ in (\ref{uM}) is to correct the
second fundamental form $h_{\alpha\beta}$ in the second parenthesis of
(\ref{CellEqNormal2}) for the change in curvature due to the normal
displacement $\psi$ -- compare (\ref{uB}), which contains this same
expression, arrived at differently.  Thus one effect of the
nonlinearity is to replace the original $h_{\alpha\beta}$ on
$\mathcal{M}$ by $h_{\alpha\beta}(\psi)$ in the balance of normal
stress.  This can be a large effect, because the tangential stress
$\sigma_\mathcal{M}^{\alpha\beta}$ can be large, and it is frequently
the $\sigma_\mathcal{M}^{\alpha\beta}h_{\alpha\beta}$ term which is
most important in balancing applied normal stress.  If
$\sigma_\mathcal{M}^{\alpha\beta}=\Sigma g^{\alpha\beta}$, for
example, corresponding to surface tension $\Sigma$ in $D_\mathcal{M}$,
then that term is the normal stress $\Sigma \Bbb{H}$, where
$\Bbb{H} = (\kappa_{1} +\kappa_{2})$ is the mean curvature.

The insight of F\"{o}ppl and von K\'arm\'an can be appreciated by
taking the special case of an initially flat $\mathcal{M}$, so that
$h_{\alpha\beta} =k_{\alpha\beta} =0$.  Then, taking Cartesian
coordinates, the covariant derivatives are ordinary derivatives, and
the equilibrium equations (\ref{CellEqTangential}) and
(\ref{CellEqNormal2}) reduce to
\begin{eqnarray}
   \label{PlateEqTangential}
   0&=&\partial_\beta\sigma_\mathcal{M}^{\alpha\beta}+f^\alpha \ ,\\
   \label{PlateEqNormal2}
   0&=&\frac{h^3}{12}(\partial_\alpha\partial_\beta
   \sigma_B^{\alpha\beta}) +h \,
   \sigma_\mathcal{M}^{\alpha\beta}(\partial_\alpha\partial_\beta\psi)
   +hf_z \ .
\end{eqnarray}
Equation (\ref{PlateEqNormal2}) is essentially the second
F\"{o}ppl-von K\'arm\'an equation.  Its middle term, representing the
normal stress due to $\sigma_\mathcal{M}^{\alpha\beta}$, is
\emph{absent} without the nonlinear term in (\ref{uM}), leaving
just bending stress to balance normal stress.  The corresponding
linear theory, which this theory was designed to correct, greatly
underestimates the strength of the membrane to resist normal stress.
As we see in (\ref{CellEqNormal2}), the contribution of the
nonlinear term persists in the case of a curved $\mathcal{M}$, but its
effect is less dramatic.

It is worth examining other aspects of the flat $\mathcal{M}$ theory
to see what else persists in the more general case.  The usual linear
relation between stress and strain, which we have not yet invoked,
should still hold in general,
\begin{equation}
\label{2dStressStrain}
\sigma^{\alpha\beta}=\frac{E}{1-\sigma^2}(g^{\alpha\mu}g^{\beta\nu}
\delta u_{\mu\nu}+\sigma\epsilon^{\alpha\mu}
\epsilon^{\beta\nu}\delta u_{\mu\nu})
\end{equation}
where $E$ is Young's modulus, $\sigma$ is Poisson's ratio, and
$\epsilon^{\alpha\beta}$ is the antisymmetric tensor with
$\epsilon^{12}=1/\sqrt{\sf g}$ and ${\sf g} = \det (g_{\alpha\beta})$.
In particular,
\begin{equation}
\label{sigmaB}
\sigma_B^{\alpha\beta}=\frac{E}{1-\sigma^2}(g^{\alpha\mu}g^{\beta\nu}
\delta u^B_{\mu\nu}+\sigma\epsilon^{\alpha\mu}
\epsilon^{\beta\nu}\delta u^B_{\mu\nu})
\end{equation}
with $\delta u^B_{\mu\nu}$ given in (\ref{uB}).  Equivalently,
strain and stress are related by
\begin{equation}
\label{2dStrainStress}
  \delta u_{\alpha\beta}=\frac{1}{E}(g_{\alpha\mu}g_{\beta\nu}
  \sigma^{\mu\nu}
   -\sigma\epsilon_{\alpha\mu} \epsilon_{\beta\nu}\sigma^{\mu\nu})
\end{equation}
and in particular
\begin{equation}
\label{2dStrainStress0} \delta
u^\mathcal{M}_{\alpha\beta}=\frac{1}{E}(g_{\alpha\mu}g_{\beta\nu}
\sigma_\mathcal{M}^{\mu\nu}-\sigma\epsilon_{\alpha\mu}
\epsilon_{\beta\nu}\sigma_\mathcal{M}^{\mu\nu}) \ .
\end{equation}
In the flat case
\begin{equation}
\partial_\beta\sigma_\mathcal{M}^{\alpha\beta}=0
\end{equation}
implies $\sigma_\mathcal{M}^{\alpha\beta}$ is derivable
\cite{airy} from an Airy potential $\chi$,
\begin{equation}
   \label{airyrep}
   \sigma_\mathcal{M}^{\alpha\beta}=\epsilon^{\alpha\mu}
   \epsilon^{\beta\nu} \partial_\mu\partial_\nu\chi
\end{equation}
eliminating one variable.  In this case, putting (\ref{airyrep})
into the right side of (\ref{2dStrainStress0}) and
$u^\mathcal{M}_{\alpha\beta}$ from (\ref{uM}) into the left side of
(\ref{2dStrainStress0}), and applying the operator
$\epsilon_{\alpha\kappa}\epsilon_{\beta\lambda}\partial_\kappa
\partial_\lambda$ to both sides, we find the first F\"{o}ppl-von
K\'arm\'an equation
\begin{equation}
\label{FvK1}
-\det(\partial_\alpha\partial_\beta\psi) =\frac{1}{E} \Delta^2 \chi
\end{equation}
and putting (\ref{airyrep}) and (\ref{sigmaB}) into
(\ref{PlateEqNormal2}) we obtain the second F\"{o}ppl-von K\'arm\'an
equation
\begin{equation}
\label{FvK2}
0= -\kappa_{c} \Delta^2 \psi
+h\epsilon^{\alpha\mu}\epsilon^{\beta\nu}
(\partial_\mu\partial_\nu\chi)(\partial_\alpha\partial_\beta\psi)+hf_z
\ .
\end{equation}
Here $\kappa_{c} = Eh^3/[12(1-\sigma^2)]$ is the bending rigidity,
with $\kappa_{c}/h \propto h^{2}$.  Tangential displacements $\delta
u^\alpha$ being small, $\det( \partial_\alpha\partial_\beta\psi)$ in
(\ref{FvK1}) equals the Gaussian curvature $\Bbb{K} = \kappa_{1}
\kappa_{2}$ of  the membrane surface $\mathcal{M}$ to fair approximation.
Bending in \emph{two} orthogonal directions, which is what bending in
general boils down to, implies \cite{mp77} $\Bbb{K} \neq 0$.

The general equations (\ref{CellEqTangential}) and 
(\ref{CellEqNormal2}) solve the long-standing problem of describing 
dynamic equilibrium of a precurved sheet under strong deformations.  
They can be widely used in shell theory and, for instance, 
to analyze crumpling \cite{lglmw95,lobkovsky,ddwvk02} of a naturally 
precurved cell membrane, such as the one in Fig.~\ref{crumpRBC}.  
Unfortunately the above program for an originally flat surface cannot 
get started in the general, curved case.  The non-commutativity of 
covariant derivatives means there is no Airy representation to 
eliminate (\ref{CellEqTangential}) even if $f^\alpha=0$.  The 
equilibrium equations, using all the tricks known for less general 
situations, remain Eqs.~(\ref{CellEqTangential}) and 
(\ref{CellEqNormal2}), with $\sigma^{\alpha\beta}$ given by 
(\ref{2dStressStrain}).  Through the second and third fundamental form 
$(h_{\alpha\beta})$ and $(k_{\alpha\beta})$ and the covariant 
derivatives they explicitly show how curvature must be taken into 
account.  Evaluating the consequences of these equations, especially 
that of taking the limit $h \to 0$, will be a true challenge to 
membrane physics for some time to come.

The authors thank Ted Steck for providing Fig.~1.  J.L.v.H. gratefully 
acknowledges constructive discussions during a stimulating workshop at 
the Aspen Center for Physics, where it all started.  M.A.P. thanks the 
SFB 563 for support while this work was done at the TU Munich.


\begin{thebibliography}{99}

\bibitem{np87}
D.R. Nelson and L. Peliti, J. Phys. France \textbf{48}, 1085 (1987).

\bibitem{sn88}
H.S. Seung, D.R. Nelson, Phys. Rev. A {\bf38}, 1005 (1988).

\bibitem{lglmw95}
A.E. Lobkovsky, S. Gentges, H. Li, D. Morse, and T.A. Witten,
Science \textbf{270}, 1482 (1995).

\bibitem{lobkovsky}
A. Lobkovsky, Phys. Rev. E {\bf 53}, 3750 (1996).

\bibitem{lw97}
A.E. Lobkovsky and T.A. Witten, Phys. Rev. E 55, 1577 (1997).

\bibitem{bap97}
M. Ben Amar and Y. Pomeau, Proc. R. Soc. Lond. A \textbf{453}, 729
(1997) and Philos. Mag. B {\bf 78}, 235 (1998).

\bibitem{kramer}
E.M. Kramer and T.A. Witten, Phys. Rev. Lett. {\bf 78}, 1303 (1997);
E.M. Kramer, J. Math. Phys. {\bf 38}, 830 (1997).

\bibitem{ccmm99}
E. Cerda, S. Cha\"{i}eb, F. Melo, and L. Mahadevan, Nature
\textbf{401}, 46 (1999).

\bibitem{audoly99}
B. Audoly, Phys. Rev. Lett. \textbf{83}, 4124 (1999).

\bibitem{bpcba00}
A. Boudaoud, P. Patricio, Y. Couder, and M. Ben Amar, Nature
\textbf{407}, 718 (2000).

\bibitem{mwwn02}
K. Matan, R.B. Williams, T.A. Witten, and S.R. Nagel, Phys. Rev.
Lett. {\bf 88}, 076101 (2002).

\bibitem{ddwvk02}
B.A. DiDonna, T.A. Witten, S.C. Venkataramani, and E.M. Kramer
Phys. Rev. E \textbf{65}, 016603 (2002).


\bibitem{bowick}
M. Bowick, A. Cacciuto, G. Thorleifsson, and A. Travesset, Phys. Rev.
Lett. {\bf87}, 148103 (2001).

\bibitem{lidmar}
J. Lidmar, L. Mirny, and D.R. Nelson, Phys. Rev. E {\bf 68}, 051910
(2003).

\bibitem{steck}
J.F. Hainfeld and T.L. Steck, J. Supramol. Struct. {\bf 6}, 301
(1977).

\bibitem{es94}
E. Sackmann,  FEBS Lett. \textbf{346}, 3 (1994).

\bibitem{npw04}
D.R. Nelson, T. Piran, and S. Weinberg, Eds., \emph{Statistical Mechanics
of Membranes and Surfaces}, 2nd ed., (World Scientific, Singapore,
2004).

\bibitem{foeppl}
A. F\"oppl, {\it Vorlesungen \"uber Technische Mechanik V} (Teubner,
Leipzig, 1907), pp.~132--144; esp.\ p.~139.

\bibitem{vonK}
Th. von K\'arm\'an, in: {\it Encyklop\"adie der Mathematischen
Wissenschaften}, Vol.~IV:4 (Teubner, Leipzig, 1910), 311--385;
esp.\ pp.\ 348--350.

\bibitem{ll59}
L.D. Landau and E.M. Lifshitz, \emph{Theory of Elasticity} (Pergamon,
London, 1959) \S14.

\bibitem{ehm64}
E.H. Mansfield, \emph{The Bending and Stretching of Plates}, 2nd ed.
(Cambridge University Press, Cambridge, 1989).

\bibitem{fluegge}
W. Fl\"ugge, {\it Stresses in Shells} (Springer, New York,
1960).

\bibitem{rdk02}
R.D. Kamien, Rev. Mod. Phys., \textbf{74}, 953 (2002).

\bibitem{vhl}
J.L. van Hemmen and C. Leibold, TU Munich Preprint.

\bibitem{hilbert} D. Hilbert and S. Cohn-Vossen, {\it Geometry and
the Imagination} (Chelsea, New York, 1952), p.~187.

\bibitem{peterson}
M.A. Peterson, J. Math. Phys. {\bf 26}, 711 (1985).

\bibitem{schutz}
B. Schutz, {\it Geometrical Methods of Mathematical Physics},
(Cambridge University Press, 1980), p.~79.

\bibitem{frankel}
T. Frankel, {\it The Geometry of Physics}, (Cambridge University
Press, 1997), p.~625.

\bibitem{airy}
G.B. Airy, Philos. Trans. R. Soc. London {\bf 153}, 49 (1863).

\bibitem{mp77}
R.S. Millman and G.D. Parker, \emph{Elements of Differential Geometry}
(Prentice-Hall, Englewood Cliffs, NJ, 1977), Sects.  4-8 \& 4-9.

\end{thebibliography}
\end{document}